# Studying the Quark Antiquark Force with Inelastic Pion Electron Scattering

J.M. LoSecco

*University of Notre Dame, Notre Dame, Indiana 46556*

(September 1 1994)

## Abstract

The concept of studying the internal structure of mesons is explored. Mesons, which are in principle two body quark antiquark interactions, may be much easier to understand than the nucleon. Measurements of the inelastic form factors to specific final states may permit careful direct studies of various components of the strong force. For example by looking at vector meson final states the spin flip amplitude can be isolated.

Technical difficulties involved in a realistic experiment are examined. Experiments to some final states such as $\rho$, $K^*$ and $a_0$ are practical today.

Subject headings: QCD — Elementary Particles — Inelastic Scattering

Typeset using REVTEX

# I. INTRODUCTION

While QCD is generally accepted as the correct theory of the strong interactions progress in understanding their "strong" nature has been slow. The great triumphs of QCD have come in the weak coupling limit. In general the theory is less tractable in the low momentum limit and there is a lack of useful data in that region too. In principle the most basic strong interaction, between a quark and an antiquark, can be best studied in the simplest bound state, mesons.

This note explores what information could be extracted from mesons with an electron probe. The creation of specific final states permits the isolation of individual components of the strong force.

To extract useful information about the 2 quark wavefunction one needs to measure the momentum transfer to the electron for specific, low multiplicity final states. For example to study the spin flip amplitude one studies the $\rho e$ or $K^* e$ final state.

# II. PHYSICS OBJECTIVES

The ideal environment to study QCD would be a system in which one had maximal control over the quark-quark interaction. Mesons provide the appropriate 2 body states but until now no probe, sensitive to only the quark degrees of freedom has been used. Electromagnetic interactions permit one to study the quark degrees of freedom. Hence the popularity of radiative decays as a probe of meson structure. But radiative decays are very limited. They give us information at a single momentum transfer . Electron probes, at low $q^2$ permit a way to study quark transitions without exciting the gluon degrees of freedom. A modest change in the quark wavefunction will not excite dramatic changes in the gluon degrees of freedom.

The interaction of an electron with an extended object can be described, in lowest order by the product of a point like interaction and a form factor which reflects the structure of the extended object [1].

$$\left(\frac{d\sigma}{dq^2}\right)_{inel} = \left(\frac{d\sigma}{dq^2}\right)_{point} \left|F_{\pi \to \rho}(q^2)\right|^2$$

The form factor, for inelastic reactions, is the Fourier transform of the product of the initial state wavefunction with the final state wavefunction. In principle, these can be calculated from the wavefunctions of the low lying resonances. Lattice gauge theories or bag models should give reasonable estimates. Traditional soft pion methods [2] may also be useful.

The inelastic form factor must vanish at $q^2 = 0$ since the two wavefunctions are orthogonal.

$$F_{\pi \to \rho}(q^2) = \int d^3 r e^{i\mathbf{q}\mathbf{r}} \psi_\rho(r) \psi_\pi(r)$$

Consider the reaction $\pi e \to \rho e$. This is simply related to the amplitude to flip the spin of one of the quarks in the pion. Similarly the reaction $\pi e \to \pi(1300) e$ is related to the amplitude to excite the dipole oscillation in the pion, into the radial excitation the $\pi(1300)$. The reaction $\pi e \to b_1(1235)e$ can be related to the amplitude to excite one unit of orbital angular momentum without spin flip. The amplitude for $\pi e \to a_0(980)e$ has both spin flip and orbital excitation contributions with spin anti parallel to the orbital angular momentum. The excitation of the $a_2(1320)$ is quite similar but the spin and orbital angular momenta are parallel.

The reaction $\pi^0 e \to \eta e$ would permit one to probe the isospin changing interaction in a very simple system. But the use of the initial state $\pi^0$ is impractical. The rare radiative decay mode $\eta \to \pi^0 \gamma\gamma$ may be of some use for this purpose.

Similar electron reactions can be identified for kaons. A mixed beam would permit one to pursue both initial states simultaneously. Comparison of pion and kaon data would give some information on mass dependence of the quark antiquark interactions.

Figure 1 plots the 4 momentum transfer squared (t) as a function of the center of mass scattering angle for pion beams of energy 600 GeV to 850 GeV incident on a stationary electron. The energy varies from 600 GeV at the top to 850 GeV at the bottom in 25 GeV steps. The figure has been drawn for the reaction $\pi e \to \rho e$. One can see that for a 600 GeV beam only $q^2 < 0.05$ (GeV/c)$^2$ are accessible. Even at 850 GeV $q^2 < 0.3$ (GeV/c)$^2$. These $q^2$ are well below what is normally considered interesting for inclusive studies.

Figure 2 illustrates a typical inelastic form factor $q^2 > 0.2$ (GeV/c)$^2$ is needed to resolve the structure of this function, that is to probe the transition in a non superficial way. Figures 1 and 2 make it clear that to truly measure the form factor one needs a beam somewhat above threshold, to span a reasonable range of momentum transfer.

The only previous experiment to study similar transitions (in the $K_L K_S$ system) [3] ran at modest energies and did not resolve the momentum transfer. It essentially only measured the charged radius of the neutral kaon. Elastic $\pi e$ scattering [4–6] has also been done at modest $q^2$ and so has not been a very deep probe of this ground state wavefunction.

### III. CROSS SECTION

For the purposes of estimating a rate let us approximate $F_{\pi \to \rho}(q^2) = A q^2 e^{-bq^2}$ which falls nicely at large $q^2$ and goes to zero at $q^2 = 0$. (This is a fit to calculations based on Hydrogen atom wavefunctions). The total cross section is finite and depends on the parameters $A$ and $b$.

$$\sigma_{total} \approx 4\pi\alpha^2 \frac{A^2}{2b}$$

Typical values of b=25. GeV$^{-2}$ and A= 7.5 GeV$^{-2}$ give a total cross section of 0.3 $\mu$barns.

### IV. KINEMATICS

Due to the low mass of the electron substantial beam energies are required to get above threshold for production of heavy mesons by electrons at rest. The natural widths of the excited states presents a range of possible beam energies. Table I indicates the pion and kaon beam momenta (in GeV/c) need to produce excited states on electrons. A beam momentum of 800 GeV/c or higher is essential to probe part of this physics. Thresholds for the $\rho$ and $K^*$ peaks are at about 560 GeV.

Note that most of these states are broad. This complicates their detection but may lower the beam energy needed to excite them. In principle one could do the experiment with the low mass portion of the $a_0$ for example. The three right hand columns give the threshold energy to excite the central value of the mass and the upper and lower edges of the state. The low center of mass energy has one positive consequence. The electrons can not excite very many degrees of freedom in the meson.

## V. HIGH ENERGIES - ELECTRON TARGET

The most direct way to excite all of the meson states considered above would be to raise the pion or kaon beam energy to $\approx 3$ TeV. No facility capable of these energies yet exists.

The obvious solution to the lack of energy in a fixed target is to have the pion interact with a stored electron beam. This not only increases the CM energy enormously but reduces, or eliminates the background from strong interactions from the fixed target protons. The major drawback of an electron target is the difficulty of achieving target brilliance (electrons/cm$^2$) comparable to a fixed target. The required beam energies are very modest by modern electron storage ring standards. For example, with a 200 GeV incident $\pi$ or $K$ beam all resonances in table I can be excited over their full widths with an electron beam of $<$ 4 MeV/c momentum (table II). Note that in the table masses are in GeV but electron beam momenta are in MeV/c.

The phase space density of the beams is always an important consideration for colliders. In this case it is anticipated that the electron beam will damp by synchrotron radiation and be fairly compact. The technical challenge is to get the meson beam to behave well. The concept of "stochastic cooling" on a particle by particle basis is possible. If the trajectories of each tagged meson can be measured, downstream transport components could be dynamically tuned to target them individually. This calls for a fairly high bandwidth feedback system and places some strong constraints on the distances between components. But it may not be impossible. The tagging rate could be limited by the bandwidth of the feedback system but one could imagine some sort of component pipeline where multiple particles are processed simultaneously.

The stored electron beam scheme would employ many of the technologies that have been developed for high brilliance synchrotron light sources and for "targeting" in linear colliders.

An advantage of an electron target is the absence of substantial hadron induced backgrounds.

## VI. CONCLUSIONS

A new probe of the quark quark interaction is potentially available via inelastic scattering from electrons. Electron excitation of low lying mesons would reveal the structure of these wavefunctions, and isolate specific QCD amplitudes. This program is an extension of the classic physics of radiative decays.

Some of these transitions could be studied today. Others require advances in accelerator technology to either higher external beam energies or a new collider technology.

## ACKNOWLEDGEMENTS

I would like to thank Neal Cason, S.U. Chung, Alex Dzierba and Nathan Isgur for help in understanding hadrons and hadron dynamics. Walter Johnson has helped to calculate inelastic form factors. Remarks by Hall Crannell about electromagnetic excitation of the giant dipole resonance in nuclei were inspirational. Hans Willutzki filled me in on details of the Mark 3 measurement of $e^+e^- \to K_L K_S$. Bill Shephard has made many helpful remarks about an earlier version of this manuscript.

FIGURES

Figure 1: t (-q$^2$) as a function of the CM scattering angle for $\pi e \to \rho e$. The different curves are for $\pi$ beam energies from 600 GeV (at top) to 850 GeV (at bottom) in 25 GeV steps.

Figure 2: Typical Inelastic Form Factor

TABLES

| State | Mass | Width ($\Gamma$) | Threshold (m) | Threshold (m+$\Gamma$) | Threshold (m-$\Gamma$) |
|---|---|---|---|---|---|
| $\rho$ | .767 | .152 | 557. | 807. | 352. |
| $a_0$ | .983 | .057 | 927. | 1040. | 820. |
| $\pi(1300)$ | 1.300 | .300 | 1636. | 2487. | 960. |
| $b_1$ | 1.232 | .155 | 1467. | 1865. | 1117. |
| $a_2$ | 1.318 | .103 | 1683. | 1960. | 1426. |
| K* | .892 | .050 | 540. | 630. | 456. |
| K*$_0$(1.4) | 1.429 | .287 | 1761. | 2645. | 1039. |
| K(1460) | 1.460 | .260 | 1849. | 2658. | 1172. |
| K$_1$(1270) | 1.270 | .090 | 1341. | 1573. | 1125. |
| K*$_2$(1.4) | 1.425 | .098 | 1751. | 2035. | 1486. |

TABLE I. Threshold Beam Energies in GeV

| State | Mass | Width ($\Gamma$) | Threshold (m) | Threshold (m+$\Gamma$) | Threshold (m-$\Gamma$) |
|---|---|---|---|---|---|
| $\rho$ | .767 | .152 | .619 | .967 | .304 |
| $a_0$ | .983 | .057 | 1.128 | 1.278 | .984 |
| $\pi(1300)$ | 1.300 | .300 | 2.057 | 3.155 | 1.172 |
| $b_1$ | 1.232 | .155 | 1.838 | 2.353 | 1.380 |
| $a_2$ | 1.318 | .103 | 2.117 | 2.475 | 1.785 |
| K* | .892 | .050 | .594 | .722 | .469 |
| K*$_0$(1.4) | 1.429 | .287 | 2.219 | 3.357 | 1.276 |
| K(1460) | 1.460 | .260 | 2.332 | 3.374 | 1.452 |
| K$_1$(1270) | 1.270 | .090 | 1.673 | 1.975 | 1.390 |
| K*$_2$(1.4) | 1.425 | .098 | 2.206 | 2.573 | 1.862 |

TABLE II. Stored Electron Momenta (MeV/c) with 200 GeV Hadron Beam

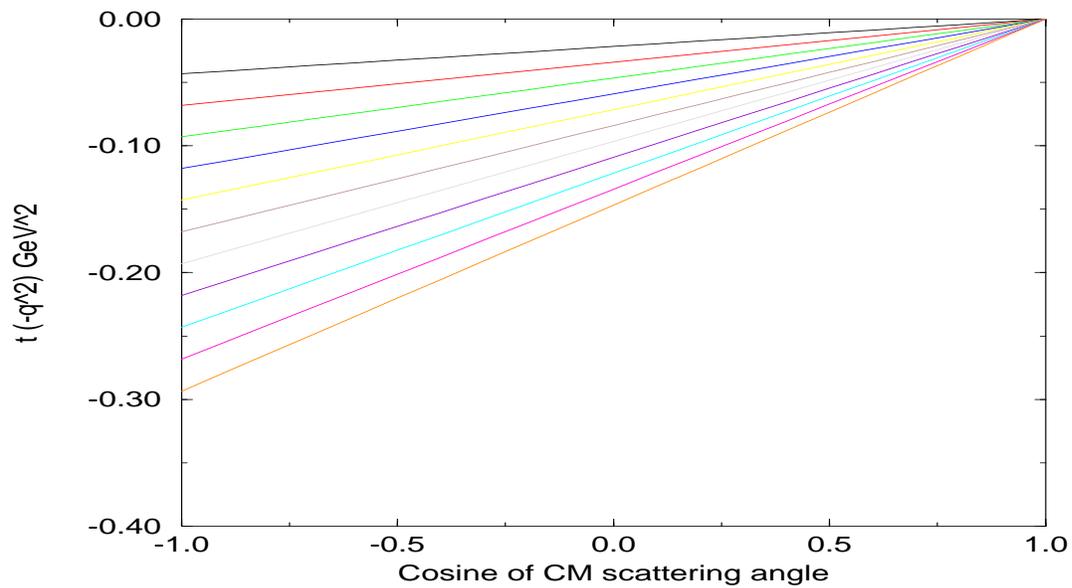

FIG. 1. t (-q²) as a function of the CM scattering angle for $\pi e \to \rho e$. The different curves are for $\pi$ beam energies from 600 GeV (at top) to 850 GeV (at bottom) in 25 GeV steps.

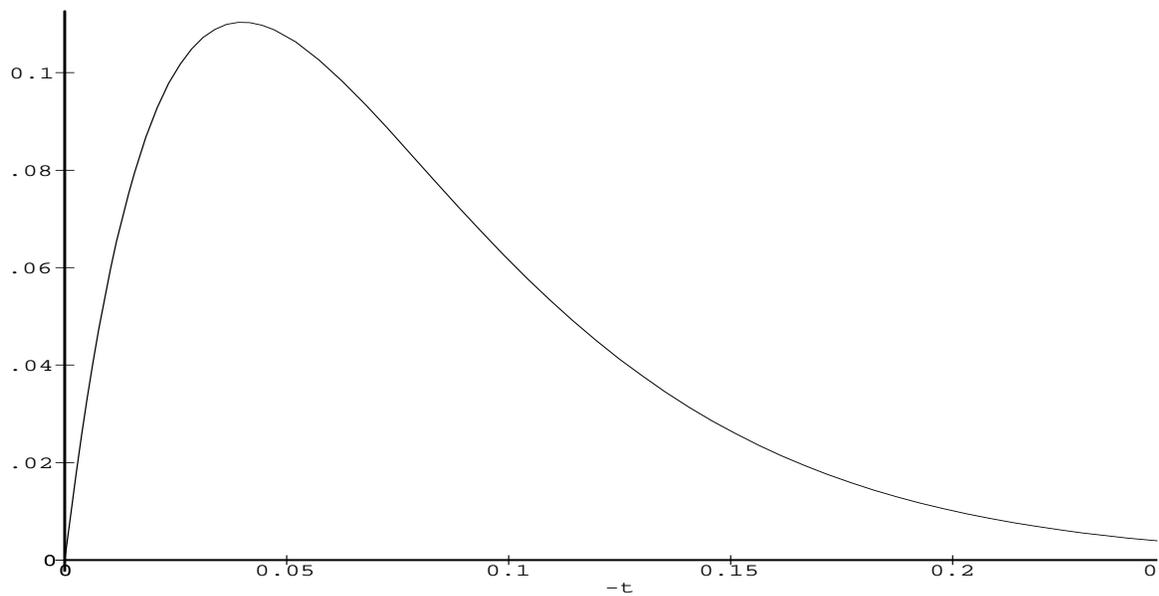

FIG. 2. Typical Inelastic Form Factor

9